\documentclass[conference]{IEEEtran}
\IEEEoverridecommandlockouts
\usepackage{amsmath}
\usepackage{graphicx}
\usepackage{cite}
\usepackage{xcolor}
\usepackage{soul}

\title{Solving Virtual Backbone Problems with Digitized Cyclic Annealing on Near-Term Quantum Computers}

\author{
  \IEEEauthorblockN{Maxence Grandadam, Maciej Koch-Janusz}
  \IEEEauthorblockA{Haiqu, Inc., 95 Third Street, San Francisco, CA 94103, USA}
}

\begin{document}
\maketitle

\begin{abstract}
Wireless multi-hop networks rely on a subset of nodes to relay traffic, broadcast control messages, and maintain global connectivity without requiring every device to forward packets. A \emph{virtual backbone} formalizes this idea by selecting a sparse set of representative nodes that can cover the network and serve as a routing substrate. In graph terms, given a general communication graph $G=(V,E)$, the backbone is often modeled as a connected dominating set (CDS): a subset $S \subseteq V$ such that every node in $V \setminus S$ has a neighbor in $S$, and the subgraph induced by $S$ is connected. CDS-based backbones reduce routing overhead but are NP-hard to compute and must balance sparsity, coverage, and connectivity. Here we introduce a quantum approach to solving the CDS problem based on the cyclic quantum annealing algorithm, suitable for current digital quantum computers, which we call Digitized Cyclic Annealing. We explore the dependence of the obtained solutions on the algorithm hyperparameters and show how transfer learning allows us to find their values for large-scale problems. We demonstrate the complete algorithm on a network optimization problem using $N=73$ qubits on an IBM Kingston quantum processor.
\end{abstract}

\begin{IEEEkeywords}
virtual backbone, maximal independent set, quantum optimization, digitized annealing
\end{IEEEkeywords}

\section{Introduction}
Formally, a set $S_{dom} \subseteq V$ is a \emph{dominating set} if $\forall v \in V$, either $v \in S_{dom}$ or $\exists u \in S_{dom}$ with $(u,v) \in E$. The \emph{minimum dominating set} (MDS) problem is to minimize $|S_{dom}|$ subject to the domination constraint. A \emph{connected} dominating set additionally requires the induced subgraph $G[S]$ to be connected, yielding the CDS problem used in backbone design.

On the other hand, an independent set is a set $S_{ind} \subseteq V$ such that $\forall u, v \in S_{ind}, \ (u, v) \notin E$. A maximal independent set (MIS) is an independent set to which no additional vertex can be added without breaking the independence constraint. Note that the largest MIS is called the \textit{Maximum} Independent Set, but is not necessarily our target. The \emph{dominating set} thus focuses on coverage, while an \emph{independent set} enforces separation among selected nodes to avoid redundant relays. Because no additional vertices can be added without creating neighboring pairs, a MIS is guaranteed to be a dominating set in any graph, and thus provides a natural backbone candidate that is sparse and well spread across the network. A connected backbone can then be obtained by adding a minimal number of Steiner nodes to connect MIS vertices \cite{Bai2020, Liang2023}, or by constructing a CDS directly with the MIS as a seed. 

This perspective supports a two-stage formulation: first, choose well-separated relay candidates via a MIS, then connect them while preserving coverage. The size of the resulting CDS depends on the size of the initial MIS used, and thus we seek a \emph{small} MIS. 
Moreover, the MIS formulation allows us to apply quantum methods to the problem. 

Concretely, we use binary variables $x_i, \ i=1 \dotsc N$ to encode which nodes belong to the independent set $S_{ind}$, and map the MIS objective to a quadratic unconstrained binary optimization (QUBO) Hamiltonian, whose ground state is the \emph{largest} MIS and each constraint violation is penalized by $U$:
\begin{equation}
    H_P = -\sum_{i \in V} x_i + U \sum_{(i, j) \in E} x_i x_j, \label{eq:Hp}
\end{equation}
This can naturally be mapped to a quantum problem by using $x_i = \frac{1+\sigma^z_i}{2}$ with $\sigma^z_i$ the Pauli Z operator for a spin-$1/2$ particle at site $i$. As discussed below this problem formulation is suitable for cyclic quantum annealing based approach, which allows us to systematically explore the low-energy solutions of the problem Hamiltonian, corresponding to MIS when the penalty term $U$ is large enough.

\section{Cyclic Quantum Annealing and Digitization} \label{ref:sec2}
After a brief overview, we introduce a digitized (that is suitable for gate-based quantum computers) quantum algorithm for the MIS problem based on cyclic quantum annealing, which we call Digitized Cyclic Annealing (DCA).
\begin{figure}
    \centering
    \includegraphics[width=.75 \linewidth]{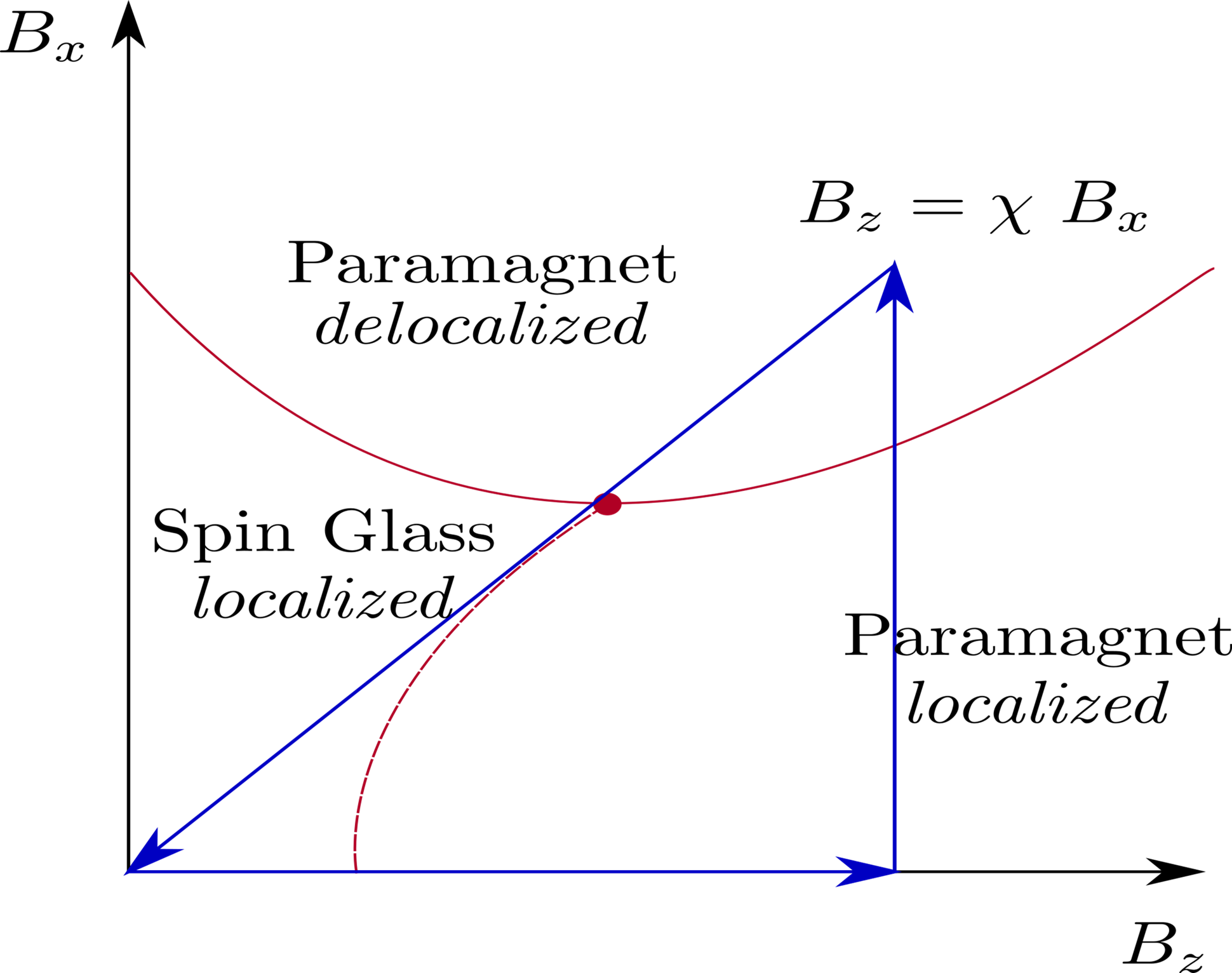}
    \caption{Schematic representation of the cyclic quantum annealing schedule. We start by ensuring that our reference state is the ground state of the classical Hamiltonian $H_P - B_z \sum_i \epsilon_i \sigma^z_i$ by turning on a longitudinal field $B_z$, effectively crossing a first order transition, depicted by the dashed red line. Then we turn on the transverse field $B_x$ to reach a delocalized phase allowing an efficient exploration of the solution space, this can be done efficiently because the gap scales as $\sqrt{B_x^2 + B_z^2}$ when crossing the second-order phase transition \cite{Laumann2012}, depicted by the solid red line. Finally, we turn off both external fields simultaneously while keeping the ratio $\chi = B_z / B_x$ fixed. If the path avoids the tri-critical point (marked by a red dot) and the first order transition line, we end up with a lower-energy state with high probability\cite{Zhang2024, Wang2022}.}
    \label{fig:Cyclic-Annealing}
\end{figure}

Quantum annealing encodes a combinatorial optimization problem into a problem Hamiltonian $H_P$ whose ground state represents the desired solution (defined for our problem in Eq.~\ref{eq:Hp}). The system is initialized in the ground state of a simple mixer Hamiltonian $H_M$ and evolved under the time-dependent Hamiltonian $H(t) = A(t) H_M + B(t) H_P$ for a total evolution time $T$ such that $A(0) \gg B(0)$ and $A(T) \ll B(T)$. If $T$ is large enough, the system stays in the instantaneous ground state at all times, and we can find the ground state of $H_P$ at the end of the annealing procedure. The exact form of $A(t)$ and $B(t)$ depends on the spectrum of the full Hamiltonian $H$, but the total evolution time needed very often scales exponentially with the system size: $T \propto e^{N}$. In practice, non-adiabatic transitions and traps in the energy landscape can lead to suboptimal solutions.

\emph{Cyclic quantum annealing} \cite{Zhang2024, Wang2022} alleviates this by repeatedly cycling the annealing schedule: starting from a reference state, we define set of biases $\epsilon_i = \langle \sigma^z_i \rangle_{ref}$ for each site, we evolve the system through a cycle in the parameter space spanned by the longitudinal field $B_z$ aligned with the biases $\epsilon_i$, and a uniform transverse field $B_x$ inducing transitions between states. The three steps of the procedure are shown in Fig.~\ref{fig:Cyclic-Annealing}: \textbf{Step 1:} the longitudinal bias field $B_z$ is gradually turned on, until the initial guess, the reference state, is the ground state of the combined Hamiltonian $H_P - B_z \sum_i \epsilon_i \sigma^z_i$. Because the Hamiltonian is diagonal in the computational ($\sigma^z$) basis, the dynamics are trivial and there are no transitions between different states. \textbf{Step 2:} the transverse field $B_x$ is gradually turned on until it reaches a fixed ratio $\chi B_x = B_z$. The field induces transitions between basis states, \emph{i.e.}~non-trivial dynamics, but the gap between the ground state and the first excited state scales as $\sqrt{B_x^2 + B_z^2}$ \cite{Laumann2012}, which allows for evolution in polynomial time. This results in the non-trivial ground state of the full Hamiltonian $H_P - \chi B_x \sum_i \epsilon_i \sigma^z_i - B_x \sum_i \sigma_i^x$. \textbf{Step 3:} both fields are gradually turned off keeping the ratio $\chi = B_z / B_x$ constant. This step is analogous to the standard annealing protocol, but we can once again perform it faster because of the polynomial nature of the gap closing, modified by the presence of the longitudinal field.

According to the adiabatic theorem, if every step is sufficiently slow, the ground state of the problem Hamiltonian is reached in one iteration. This is similar to the standard quantum annealing, but the total evolution time now scales polynomially, and not exponentially in the system size. However, for large systems, with many local minima, this is still not feasible in practice. Furthermore, near-term devices introduce additionally quantum noise, perturbing the ideal evolution.

The key advantage of cyclic annealing is that if the ratio between the longitudinal and transverse fields is chosen well (see Fig.~\ref{fig:Cyclic-Annealing}), at the end of each cycle with high probability \cite{Zhang2024, Wang2022} a lower energy state is obtained than the initial reference, which allows us to iteratively improve the solution. Moreover, this allows us to sample
low-energy solutions, potentially more useful than the exact ground state of the problem Hamiltonian \cite{Pennington2025}, such as in our CDS construction, where different small MIS (\emph{i.e.}~not the \emph{maximum} one) can be used as seeds \cite{Bai2020, Liang2023}.

As mentioned, the key effect of the longitudinal field is to induce a polynomial scaling of the gap between the ground state and the first excited state \cite{Wang2022}, allowing to avoid the exponential slowdown typically encountered in annealing. This makes execution on near-term QPUs feasible, but the annealing is considerably more complex, with three distinct steps and schedules. It can, however, be made simpler: in the regime where the ratio of the fields $B_x$ and $B_z$ to the problem Hamiltonian is large we can directly prepare the ground state expected after steps 1 and 2 on a digital quantum computer, as it is a product state. Step 3, reducing the fields while keeping $\chi = B_z / B_x$ constant, is performed using a quantum alternating-operator ansatz (QAOA) \cite{Farhi2014}, similarly as in ``warm-start QAOA" \cite{Egger2021, Romero2025, Lopez-Ruiz2025}. We refer to the above digital implementation of the cyclic quantum annealing protocol as \emph{Digitized Cyclic Annealing} in the remainder of the paper.

An important part of the algorithm is the implementation of time evolution under the problem and mixer Hamiltonians. On a digital quantum computer, the evolution under the problem Hamiltonian can be decomposed into $ZZ$ rotations, then further decomposed into elementary two-qubit gates depending on the target hardware. The mixer Hamiltonian is modified compared to the standard QAOA algorithm and given by:
\begin{equation}
    H_M = -\sum_{i \in V} \left( \sin(\theta_i) \sigma^x_i + \cos(\theta_i) \sigma^z_i \right), \label{eq:Hm}
\end{equation}
where $\theta_i = \pi/2-\epsilon_i \arctan(\chi)$, which recovers the usual X-mixer for $\chi \to 0$. Note that the sign in front of the longitudinal field changes when $\epsilon_i$ changes, biasing the system toward the initial guess, while the transverse field remains uniform.
Despite containing two non-commuting terms, the evolution under the mixer Hamiltonian can also be implemented exactly:
\begin{equation}
    e^{-it \left( -\sin(\theta) \sigma^x - \cos(\theta) \sigma^z \right)} = \hat{R}_Y\left( \theta \right) \hat{R}_Z\left( -2t \right) \hat{R}_Y\left( -\theta \right), 
\end{equation}
where $\hat{R}_a$ is a single-qubit rotation around the $a$-axis (using the Qiskit rotation angle convention). The total cyclic annealing evolution can thus be approximated on a digital quantum computer using a $p$-layer QAOA, with the problem and mixer Hamiltonians given by Eqs.~\ref{eq:Hp}--\ref{eq:Hm}.

To faithfully execute the cyclic annealing protocol, we need to efficiently prepare the ground state of the mixer Hamiltonian. Since it is entirely local, the ground state can be derived and implemented using only single-qubit rotations. For site $i$, a single $\theta_i$ rotation around the $Y$ axis is sufficient:
\begin{equation}
    \lvert \phi_0 \rangle_i = \hat{R}_Y\left(\theta_i \right) \lvert 0 \rangle,
\end{equation}
which again recovers the uniform superposition over all basis states used by the usual mixer for $\chi \to 0$.
\begin{figure}[t]
    \centering
    \includegraphics[width=\linewidth]{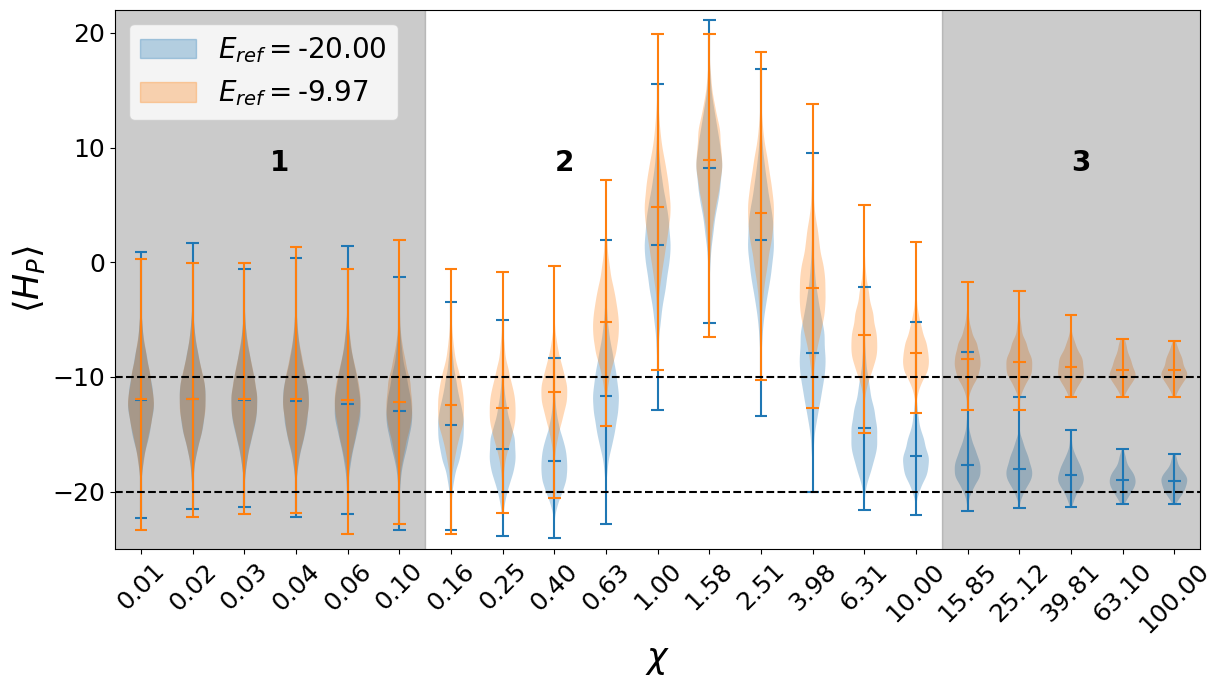}
    \caption{Energy distributions of the states obtained when sampling from a DCA circuit of depth $p=10$ for $N=100$ qubits, and for two different reference states (dashed black lines). We observe three distinct regimes: A reference-independent result for small $\chi$ (region 1, shaded), with overlapping distributions regardless of the reference state used. The localization transition for $\chi \sim 1.6$ (region 2), where the average energy reaches a minimum before increasing rapidly when we cross the tricritical point. Lastly, a concentration around the reference state for large $\chi$ (region 3, shaded), with a much smaller variance in the energy of the measured states. This phenomenon is discussed in more detail in the main text. The test problem is a 1D antiferromagnetic Ising model encoding a simple example of the MIS problem. Each distribution was obtained by sampling $n_{shots} = 10\,000$ shots on a  Matrix Product State (MPS) simulator.}
    \label{fig:localization-simulation}
\end{figure}

In the original QAOA\cite{Farhi2014}, the time-step durations of the alternating evolutions are determined variationally to minimize the expectation value of the Hamiltonian at the end. Recently, it has been suggested that training such an Ansatz on noisy QPUs is difficult, both theoretically \cite{Bittel2021} and in practice \cite{Schumann2024}. Different approaches have been proposed to replace variational training with classical methods approximating the optimal schedule, such as transfer learning \cite{Bach2025, MontaezBarrera2025a}, Fourier representations \cite{Zhou2020}, or transfer-state methods \cite{Sack2023}. In our experiments, we use the ``linear-ramp QAOA" (LR-QAOA) \cite{MontaezBarrera2025b}, a heuristic approach based on quantum annealing intuitions, as it requires few parameters and has been shown to be robust if the number of QAOA layers scales with the problem size.

Our DCA circuits for each step are thus given by
\begin{equation}
    U\left( \Delta, \chi \right) = \prod_{k=0}^{p-1} e^{-i \beta_k H_M\left( \chi \right)} e^{-i \gamma_k H_P}
\end{equation}
with the time-step durations $\beta_k$ and $\gamma_k$ given by
\begin{equation}
    \beta_k = \left( 1-\frac{k}{p} \right) \Delta \qquad \gamma_k = \frac{k+1}{p} \Delta,
\end{equation}
where $\Delta$ and $\chi$ control the dynamics introduced by the mixer, and the bias introduced by the longitudinal field, respectively.

\section{Results on Digitized Cyclic Quantum Annealing}\label{ref:sec3}
\subsection{Study of the localization transition}
Here we study the dependence of  the distribution of the states obtained at the end of DCA on the longitudinal field $B_z$. We use a simple 1D antiferromagnetic Ising system as a model to allow efficient simulations of the associated quantum circuits. The problem Hamiltonian is thus given by
\begin{equation}
    H_P = J \sum_{i=0}^{N-2} \sigma^z_i \sigma^z_{i+1}
\end{equation}

The results of the DCA protocol are shown in Fig.~\ref{fig:localization-simulation} for two different reference states and different values of $\chi$ (or, equivalently, of $B_z$). Three regimes can be distinguished: \textbf{Region 1:} For small $\chi$, we recover the standard LR-QAOA behavior: the distribution is independent of the reference state (and, as in QAOA, will depend on the number of layers $p$, the schedule, and its relation to the Hamiltonian).
\textbf{Region 2:} Close to the tricritical point, we observe a minimum in the average energy of the measured states due to the efficient exploration of the solution space induced by the longitudinal field. \textbf{Region 3:} For large $\chi$, the distribution concentrates on the reference state, and thus even if the average energy of the sampled solutions is small, the probability of improving upon the reference state decreases rapidly. This is the localized regime.
\begin{figure}[t]
    \centering
    \includegraphics[width=\linewidth]{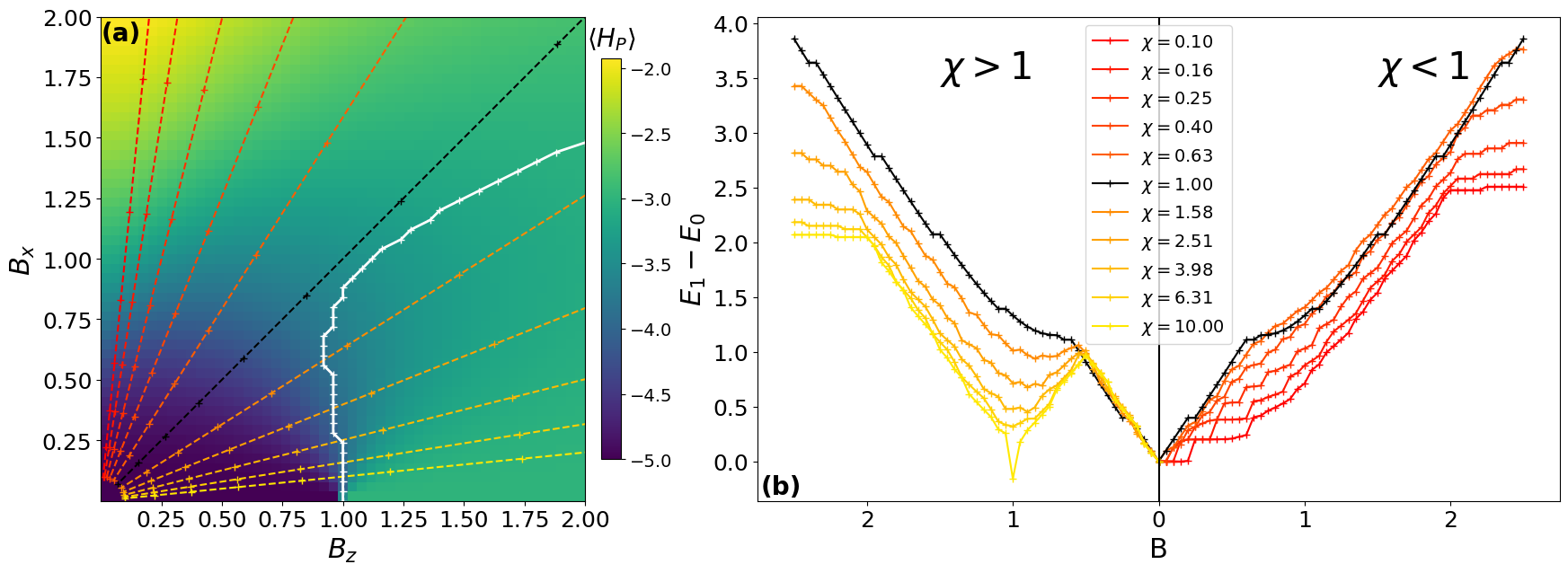}
    \caption{$\textbf{(a)}$ Phase diagram of a small $N=7$ antiferromagnetic Ising model on a 1D chain with both longitudinal $B_z$ and transverse $B_x$ fields. We identify the first-order phase transition (white line) between the frustrated ground state of the problem Hamiltonian at $\left(B_z=0, B_x=0\right)$ and the reference state at $\left(B_z>B_z^*, B_x=0\right)$ by the discontinuity in the average energy of the problem Hamiltonian. This transition is extended to $B_x \neq 0$ by tracking the maximum of the derivative of this expectation value. The second-order transition between the localized spin-glass and the delocalized paramagnet is not visible in the example due to the small system size. Dashed colored lines correspond to the paths $B_z = \chi \ B_x$ used in panel (b) with different values of $\chi$. $\textbf{(b)}$ Energy difference between the first excited state ($E_1$) and the ground state ($E_0$) for different choices of $\chi$. We plot paths crossing the first-order transition ($\chi \gg 1$) and paths avoiding it ($\chi \ll 1$). The black line marks $\chi=1$, appearing on both sides for easier comparison. We clearly observe two gap closings for large values of $\chi$.}
    \label{fig:spectrum}
\end{figure}

The behavior of the average energy in Regime 2, which attains a minimum before rising rapidly, is a signature of the localization transition. This supports our  schedule choice, which aims to be as close as possible to the tricritical point in Fig.~\ref{fig:Cyclic-Annealing} without crossing the first-order transition.

The identification of the sudden increase in average energy with the localization transition can be made more precise by examining the spectrum of the Hamiltonian for different values of $B_x$ and $B_z$. We use a smaller system with $N=7$ sites to allow for exact diagonalization on a dense grid of $\left( B_z, B_x \right)$ values. The average energy with respect to the problem Hamiltonian $\langle H_P \rangle$ is shown in Fig.~\ref{fig:spectrum}(a), and the localization transition at $B_x=0$ is easily identified as the discontinuous jump in $\langle H_P \rangle$, when the reference state becomes the ground state of the classical Hamiltonian $H_P - B_z \sum_i \epsilon_i \sigma^z_i$. The transition can be extended to $B_x \neq 0$ by following this discontinuity. Because of the limited system size and the low amount of frustration in the 1D system we use, the transition between the spin-glass phase and the delocalized phase is not as well defined. We thus only observe the first-order transition (white line) between the delocalized phase and the localized paramagnet.

In Fig.~\ref{fig:spectrum}(b), we plot the gap between the ground state and the first excited state when reducing $B_z = \chi B$ for different values of $\chi$. The corresponding trajectories in the phase diagram are depicted in Fig.~\ref{fig:spectrum}(a) using correspondingly colored lines. As expected, the gap is larger in the presence of a longitudinal field ($\chi \neq 0$), allowing for a faster adiabatic schedule. However, we also observe that when crossing the localization transition, we encounter a second gap closing around $B \sim 1$. This is the origin of the large increase in the average energy of the states measured in our cyclic annealing protocol, as this gap closing leads to Landau-Zener transitions to higher-energy states. Note that for large $\chi \gg 1$ the gap closing still occurs during the annealing procedure, but because the total Hamiltonian is almost diagonal ($B_x = B_z / \chi \ll 1$), the dynamics are almost trivial and the transition probabilities are suppressed. A complete characterization of the two transitions would require a genuinely frustrated system of larger size, but we expect these properties to hold.
\begin{figure}[t]
    \centering
    \includegraphics[width=\linewidth]{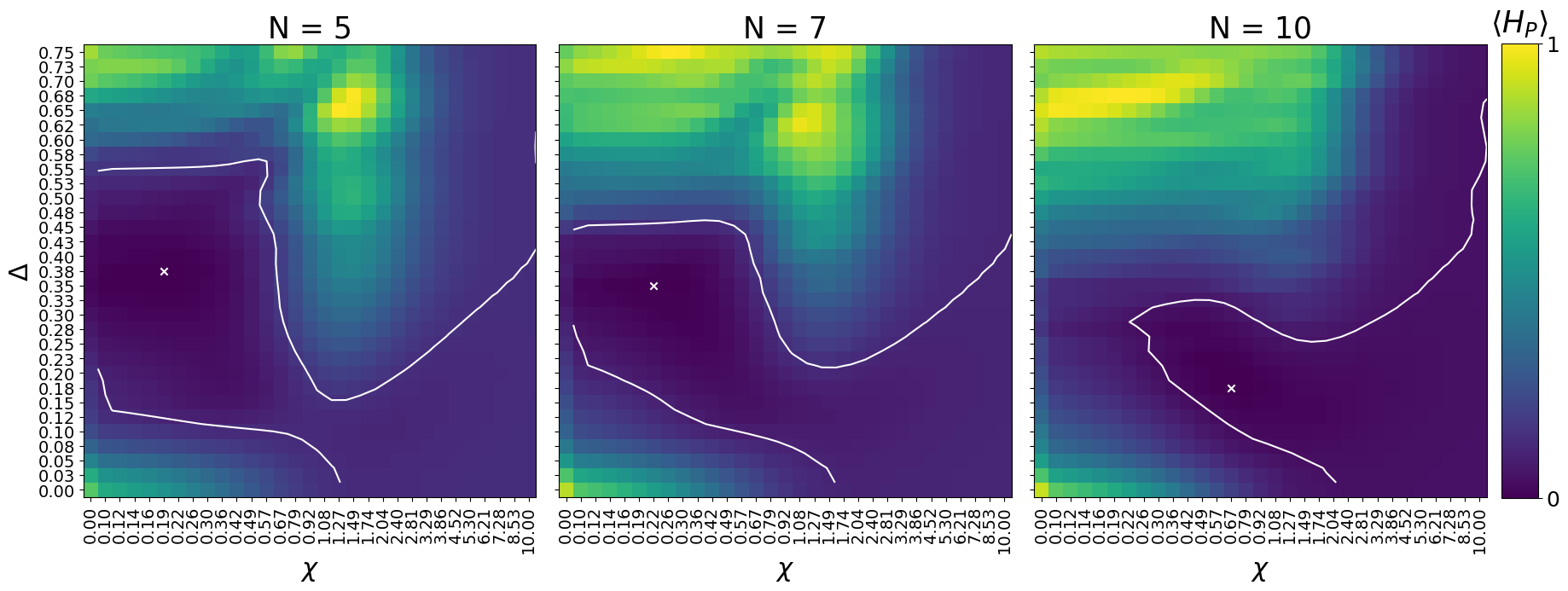}
    \caption{The average energy  obtained for three random unit-disk graphs of different sizes $N$ using a DCA circuit with $p=10$ when varying $\Delta$ and $\chi$. White cross marks the minimum expectation value for each case, while the white line delimits the region where the average energy is lower than that of the reference state. The average energy has been normalized to allow a comparison between the different system sizes.}
    \label{fig:Grid_search}
\end{figure}

\section{Virtual Backbone on 73 Qubits}\label{ref:sec4}
We now implement our iterative procedure to solve the minimum dominating set problem for a random unit-disk graph with $N=73$ nodes. The graph is not tied to any particular layout of a quantum processor (QPU). While the DCA circuit would require significant classical resources to simulate, the optimization problem itself remains easily solvable with standard classical solvers such as CPLEX.

\subsection{Transfer learning to fix $\Delta$ and $\chi$}
To achieve optimal performance with our iterative solver, we need to choose $\Delta$ and $\chi$ according to the spectrum of the time-dependent Hamiltonian. Since finding them directly for large problems is difficult in general, we turn to transfer learning, where parameters determined from related smaller instances are used to fix those for the instance of interest.

Here, we determine the optimal $\Delta$ and $\chi$ by a grid-search. We use MPS to simulate the $p=10$ DCA for $20$ MIS problems on random unit-disk graphs of sizes $N=5,\ 6,\ 7,\ 8,\ 9,\ 10$ with similar characteristics, over a grid of $\Delta$ and $\chi$. We then extrapolate the parameters to use for the $N=73$ instance. This approach has been  applied to the standard LR-QAOA without a longitudinal field \cite{Dehn2026}. We extract both the standard parameter set $(\Delta^{\prime\prime}, \chi=0)$ for LR-QAOA and the optimal set $(\Delta^{\prime}, \chi^*)$ allowing for a non-zero longitudinal field.

\begin{figure}
    \centering
    \includegraphics[width=\linewidth]{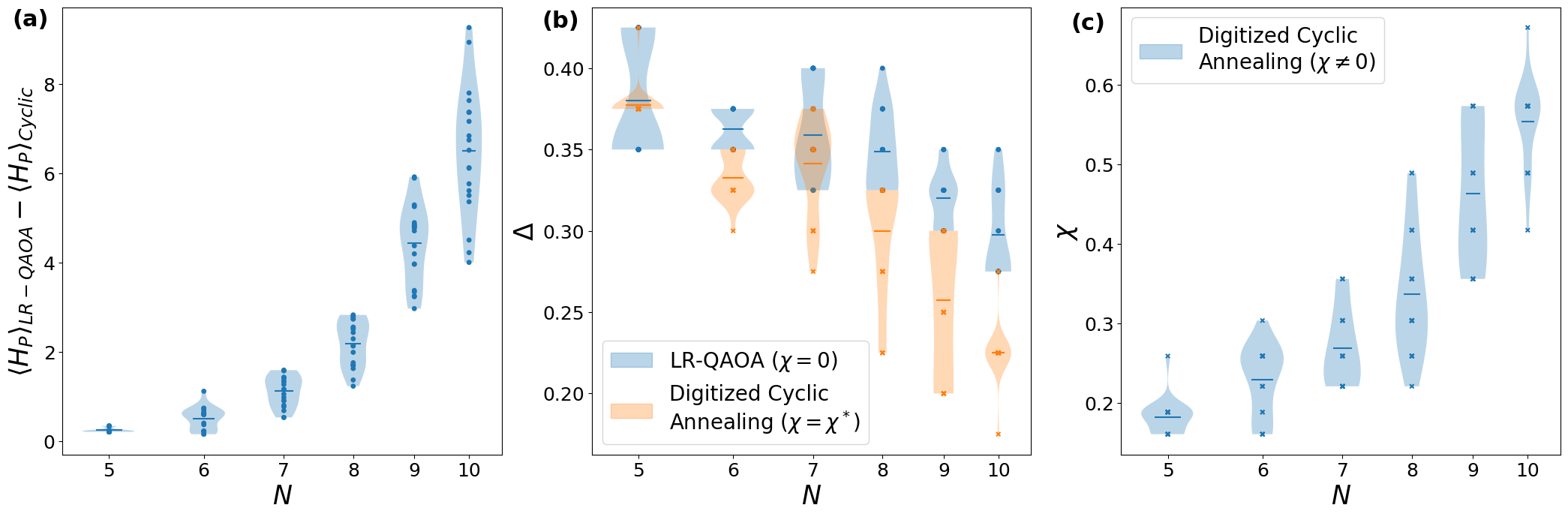}
    \caption{\textbf{(a)} Difference between the best expectation values obtained from standard LR-QAOA $(\Delta^{\prime\prime}, \chi=0)$ and DCA $(\Delta^{\prime}, \chi^*)$ with $p=10$. DCA consistently outperforms the standard LR-QAOA. The advantage grows with the system size, as expected when keeping the number of QAOA layers $p$ constant. \textbf{(b)} Values of $\Delta$ leading to the optimal performance with and without a longitudinal field. \textbf{(c)} Values of $\chi$ leading to the optimal performance for DCA with $p=10$ for different graph sizes. These are used to extrapolate the optimal parameters to $N=73$.}
    \label{fig:transfer_learning}
\end{figure}

In Fig.~\ref{fig:Grid_search} we show the grid-search results. The optimization landscape is robust across different problem sizes, and random problem instances. 
The optimal point consistently occurs at a non-zero value $\chi = \chi^* \neq 0$, indicating that this approach is beneficial even for small problems. We compare the best solutions obtained from standard LR-QAOA ($\chi=0$) and DCA in Fig.~\ref{fig:transfer_learning}(a). We observe a consistent improvement in the final expectation value for all instances when allowing for $\chi \neq 0$. This improvement increases with graph size, which is expected since the ratio $p/N$ decreases as $N$ grows while $p$ is fixed. We also observe in Fig.~\ref{fig:transfer_learning}(b) that the optimal value $\Delta^{\prime}$ shifts when $\chi \neq 0$, indicating a non-trivial modification of the optimal schedule. Furthermore, Fig.~\ref{fig:transfer_learning}(c) shows that $\chi$ evolves smoothly with the system size, supporting the validity of the transfer learning approach. We use these results to fit the dependence of $\Delta$ and $\chi$ on the graph size.

\subsection{Solving a large instance on a Quantum Computer}
\begin{table}[h!]
\centering
\begin{tabular}{ |p{.3 \linewidth}||p{.25 \linewidth}|p{.25 \linewidth}|  }
 \hline
 \multicolumn{3}{|c|}{Circuit statistics} \\
 \hline
    & Two-qubit count & Two-qubit depth \\
 \hline
 Qiskit transpilation   & 6648    & 980 \\
  \hline
 Haiqu transpilation &   2002  & 335 \\
 \hline
\end{tabular}
\newline
\caption{}
\label{table:circuit_statistic}
\end{table}

We execute the $N=73$ DCA circuits on the IBM Kingston quantum processor. The necessary circuit transpilation (using the Qiskit transpiler \cite{Javadi-Abhari2024}) to the device increases, however, the two-qubit gate count and depth far beyond the capabilities of the processor (see Table~\ref{table:circuit_statistic}). We therefore apply two circuit optimization techniques to improve performance.

We first apply a hardware-aware pruning technique that removes long-range gates requiring many SWAP operations on the device topology \cite{Escofet2026}. A gate is pruned if the fidelity loss from its removal (dependent on its rotation angle) is smaller than the expected fidelity loss from the additional noisy SWAP operations required to implement it. We then further compress the circuit using Haiqu's tensor-network-based approximate compiling. The final gate counts obtained are shown under ``Haiqu transpilation'' in Table~\ref{table:circuit_statistic}.

\begin{figure}
    \centering
    \includegraphics[width=\linewidth]{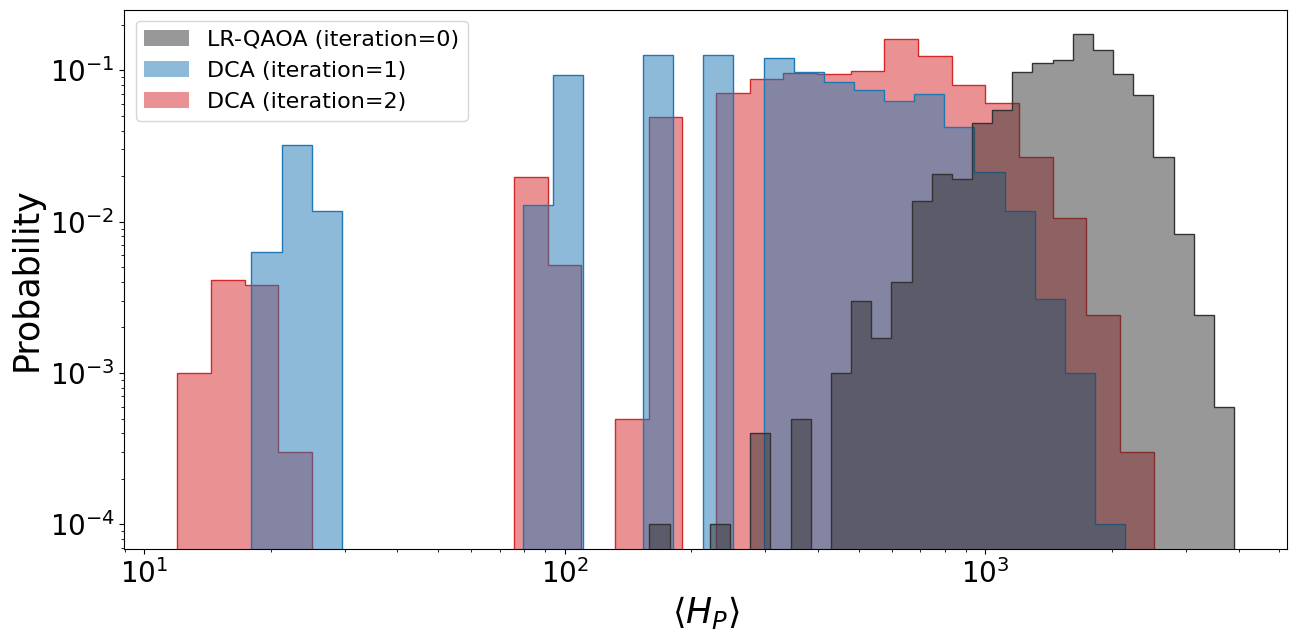}
    \caption{Results from iterative DCA on quantum hardware. The first iteration uses standard LR-QAOA ($\chi=0$), and each subsequent iteration uses the best solution from the previous one as a reference. Despite hardware noise, the sampling of low-energy states improves across iterations. Experiments were run on the IBM Kingston device with $10\,000$ shots per iteration.}
    \label{fig:QPU_results}
\end{figure}

With the obtained circuit we run three consecutive iterations of DCA on the quantum processor, and the resulting energy distributions are shown in Fig.~\ref{fig:QPU_results}. In the first iteration, we use $\chi=0$ to obtain an unbiased starting point, equivalent to the standard QAOA. The first-iteration results improve over random sampling and yield valid independent sets, although none are maximal independent sets.

Using the best solution from the first iteration as a reference, the second iteration yields a significantly improved distribution. While many sampled bitstrings are worse than the reference state due to hardware noise, a substantial fraction improves upon it. This behavior is common for QAOA-type algorithms on current noisy hardware and can, in theory, be circumvented by increasing the number of samples taken from the quantum computer by a factor proportional to the noise strength \cite{Barron2024}. This improvement continues in the third iteration, demonstrating that the method can reliably outperform standard LR-QAOA and efficiently explore the low-energy subspace of the MIS Hamiltonian even on current noisy QPUs.

Note that we did not find the ground state of the problem Hamiltonian on the noisy QPU, but this was not the main goal of the quantum procedure. In fact, as mentioned earlier, the ground state of the problem Hamiltonian $H_P$ would be the largest (\emph{maximum}) MIS, which is not the best seed for obtaining the smallest connected dominating set. Instead, the DCA algorithm allows us to tune how localized, in Hamming distance with respect to the reference state, the exploration for new solutions is. In our example, we see that $\sim 5 \%$ of the samples we obtain from the DCA circuit on QPU are valid independent sets.

\subsection{From independent sets to connected dominating sets}

Here we complete the virtual backbone construction for our $N=73$ graph example, by adding to the MIS obtained from the quantum samples a minimal number of Steiner nodes connecting them. We obtain a CDS in this fashion\cite{Bai2020, Liang2023}.
\begin{figure}
    \centering
    \includegraphics[width=\linewidth]{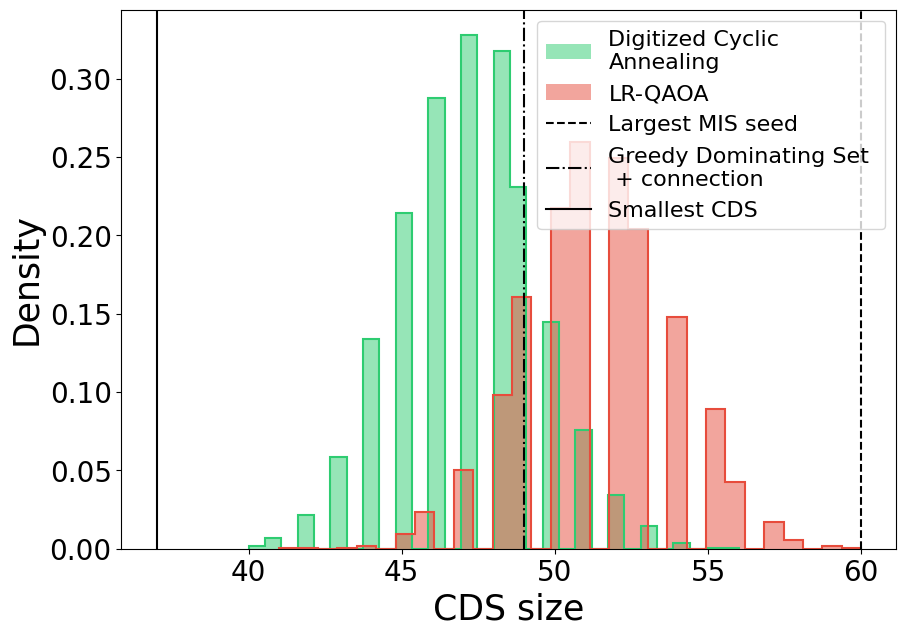}
    \caption{Results from the CDS construction for the $N=73$ graph based on the samples obtained from either the LR-QAOA circuit or the DCA circuit on the IBM quantum processor. The vertical lines mark different classical baselines: exact minimum CDS (solid), standard greedy dominating set construction + connection (dashed dotted) and CDS obtained from the largest MIS of the underlying graph (dashed). The DCA results provide better CDS than the standard LR-QAOA or than a classical greedy construction on average, but do not find the minimum CDS for this example.}
    \label{fig:CDS}
\end{figure}

The size distribution of CDS obtained by sampling on the QPU from the LR-QAOA and DCA circuits is shown in Fig.~\ref{fig:CDS}. We compare it to three classical baselines: the minimum CDS found using a standard integer programming solver has $n_{\text{MCDS}} = 37$ nodes. A standard greedy classical technique to create CDS is to add nodes with the largest degree until the graph is dominated. This dominating set is then connected to obtain a CDS. Applying this method to our graph yields a CDS with $n_{\text{greedy}} = 49$ nodes. We also include the CDS obtained from the largest maximal independent set for our graph, \emph{i.e.}~the CDS obtained from the ground-state of $H_P$. As expected, the result is larger with $n_{\text{LMIS}} = 60$ nodes.

As shown in Fig.~\ref{fig:CDS} the average size of the CDS obtained from our DCA solutions is smaller than the average size obtained using LR-QAOA. Moreover, even on the noisy quantum processor, the results are comparable with those achieved by the classical greedy strategy. As remarked before, being able to explore low-energy solutions of $H_P$ without necessarily targeting the ground state is an asset, as the CDS obtained from the largest (\emph{maximum}) MIS is much larger.

\section{Conclusion and Extensions}

In this work, we introduced a Digitized Cyclic Annealing framework for solving virtual backbone construction problems in wireless communication networks. By formulating the problem through a maximal independent set (MIS) encoding and mapping it to a QUBO Hamiltonian, we demonstrated how quantum optimization techniques can be effectively leveraged to address a fundamental network design challenge.

Our results show that cyclic annealing significantly improves upon standard QAOA-based approaches by enhancing the exploration of low-energy configurations through the controlled use of longitudinal fields. In particular, we observed consistent gains in solution quality compared to the standard QAOA algorithm, with improved sampling of valid independent sets and progressive refinement through iterative execution on quantum hardware. The control offered by tuning the localization strength through the field-ratio parameter $\chi$ is also useful when there are valid solutions in the low-energy sector of the problem Hamiltonian, which are not necessarily the ground state, as in our example. The experimental validation on IBM quantum processors further highlights the practical feasibility of the approach under NISQ (noisy intermediate-scale quantum) conditions, especially when combined with hardware-aware circuit optimization and compression.

Moreover, the use of transfer learning to determine the cyclic annealing parameters for different graph sizes demonstrates the potential for reducing costly variational overhead while preserving performance. The additional grounding of the hyperparameters in the physics of the system makes the approach more interpretable than other iterative variants of quantum annealing. This positions cyclic annealing as a promising building block for emerging quantum machine learning frameworks applied to communication networks.

Overall, our findings indicate that hybrid quantum-classical approaches based on cyclic annealing can provide a scalable and adaptable pathway toward efficient backbone construction and, more broadly, toward intelligent network optimization. Several extensions are natural. First, larger and denser network instances should be tested to assess scalability and robustness. Second, hybrid classical-quantum workflows, such as initializing with polynomial-time approximations or applying simulated annealing post-processing, may further improve performance. Finally, alternative sampling strategies such as \emph{parallel tempering} \cite{Earl2005} provide complementary mechanisms for exploring rugged energy landscapes by combining local search and large moves, and could find a natural extension with cyclic annealing via tuning of $\chi$.

\bibliographystyle{IEEEtran}
\bibliography{References}

\end{document}